# Probing disorder-induced time-reversal symmetry breaking in Josephson junctions

Yu Wu[1,2,*], Daiqiang Huang[3,*], Huanyu Zhang[1,2], Anita Guarino[4], Rosalba Fittipaldi[4], Chao Ma[5], Wenjie Hu[1,2], Chang Niu[1,2], Zhen Wang[6], Weichao Yu[1,7], Yuriy Yerin[8], Antonio Vecchione[4], Yang Liu[3], Mario Cuoco[4,a)], Hangwen Guo[1,7,9,a)], and Jian Shen[1,2,7,9,10,a)]

[1]State Key Laboratory of Surface Physics and Institute for Nanoelectronic Devices and Quantum Computing, Fudan University, Shanghai 200433, China

[2]Department of Physics, Fudan University, Shanghai, 200433, China

[3]International Center for Quantum Materials, Peking University, Beijing, 100871, China

[4]CNR-SPIN, c/o Universit´a di Salerno, I-84084 Fisciano (SA), Italy

[5]College of Materials Science and Engineering, Hunan University, Changsha 410082, China

[6]Department of Physics, University of Science and Technology of China, Hefei, Anhui 230026, China

[7]Zhangjiang Fudan International Innovation Center, Fudan University, Shanghai 201210, China

[8]CNR-SPIN, via del Fosso del Cavaliere, 100, 00133 Roma, Italy

[9]Shanghai Research Center for Quantum Sciences, Shanghai 201315, China

[10]Collaborative Innovation Center of Advanced Microstructures, Nanjing 210093, China

a) email: #mario.cuoco@spin.cnr.it †hangwenguo@fudan.edu.cn ‡shenj5494@fudan.edu.cn

**Abstract**

The relation between superconductivity and time-reversal symmetry (TRS) is one of the most fascinating problems in condensed matter physics. Although most superconductors inherently possess TRS, nonmagnetic disorder can induce states that demonstrate the breaking of this symmetry. Yet, the identification of experimental signatures of superconductivity with broken TRS remains a challenge. Here, we fabricate vertical Josephson junctions using metallic superconductor (Al) and ion bombarded $Sr_2RuO_4$ to study disorder-driven TRS breaking effects. We observe persistent magneto-resistive hysteresis behavior dependent on the disorder deposition time that provides evidence of TRS breaking below the superconducting transition temperature. Field and temperature dependent measurements suggest that the observed effects arise from disorder-induced anomalous flux in $Sr_2RuO_4$ surface which can be sensitively detected by superconducting Al. Our experimental results can be accounted within a physical framework of disorder-induced reconstruction of the superconducting order parameter as described within a multiband Ginzburg-Landau approach.

# Introduction

Time-reversal symmetry breaking (TRSB) phenomena in superconductors are among the most interesting and debating problems in condensed matter physics. To date, while several superconductors[1-6] have shown experimental findings that can be consistent with TRSB, detection and origin of the emergent phases remain difficult and controversial[7-9]. Typically in superconductors with TRSB the occurrence of internal magnetic fields is either due to the magnetic moment of the Cooper pairs, as in non-unitary spin-triplet pairing, or by the multicomponent nature of the superconducting state. For the latter, it is the complex superposition of distinct order parameters that leads to a breaking of time-reversal symmetry. In this context, it is known that disorder can cooperate to favor the formation of time-reversal symmetry broken phases[10-21]. On the other hand, experimental signatures on the role of disorder to induce TRSB states and phenomena in superconductors remain a challenge[22-25]. In this context it would be highly desirable to have signatures of TRSB effects by exploiting electronic transport measurements. Owing to the zero-resistance superconducting background, identifying TRSB requires an electronic probe that is sensitive to the anomaly of local flux or domain variations, thus calling for a careful design of materials involved and of the measurement geometries.

In this Letter, we fabricate Josephson junctions consisting of $Sr_2RuO_4$ superconductor and metallic superconductor Aluminum (Al) to explore disorder-induced TRSB effect. $Sr_2RuO_4$ is an unconventional superconductor whose origin and pairing structure remain a mystery[26,27]. There have been evidences of the occurrence of time-reversal symmetry breaking in the bulk of $Sr_2RuO_4$ from zero-field muon spin spectroscopy[28,29] and Kerr effect measurements[30]. Similar conclusions have been achieved by Josephson effects[31,32] and SQUID phase sensitive experiments[33]. The fact that surface magnetic signatures[34-36] have been observed above the superconducting temperature indicates that there might be an enhancement of electron correlations in those regions of the sample. For this reason, it is relevant to assess whether nearby the surface there might occur an anomalous reconstruction of the superconducting state. To this aim, nonmagnetic disorder is introduced on the surface of $Sr_2RuO_4$ by controllable $Ar^+$ bombardment[37,38]. Via transport measurement, we observed distinct and reproducible magneto-resistive hysteresis behavior that are signatures of TRSB effects in the superconducting state. Our experimental observations can be ascribed to local flux trapping in $Sr_2RuO_4$ due to disorder and magnetic field driven superconducting phase reorganization of the order parameter. Such flux anomaly is detectable thanks to the small critical field of Al, which exhibit high spatial sensitivity to the local flux distribution in the $Sr_2RuO_4$ layer.

## Results

To study the effect of disorder on TRSB, we fabricated the Josephson junctions by the following procedures. As shown in Figure 1a, high quality single crystal $Sr_2RuO_4$ (supplementary Figure S1[39]) was cleaved in glove box and covered by mask with open size of 200 um×200 um. To introduce disorder, the sample is transferred in-situ for low energy $Ar^+$ bombardment on $Sr_2RuO_4$ surface. The bombarded $Sr_2RuO_4$ was then transferred in-situ into the sputtering chamber and 400 nm thick Al films were grown to form Josephson junctions. Figure 1b shows an example of the cross-sectional atomic-resolution high-angle annular dark-field (HAADF) STEM image of the junction with 100 min $Ar^+$ bombardment time. At the interface between the $Sr_2RuO_4$ with a complete lattice structure and amorphous Al layer, a region containing spatially inhomogeneous $Sr_2RuO_4$ nanoclusters can be observed. We employed a three-terminal method[40] which is sensitive to the interfacial region of the Josephson junction to conduct transport measurement (Figure 1a). Figure 1c shows the resistivity vs temperature curve of this junction. With decreasing temperature, the superconducting transition of $Sr_2RuO_4$ and Al are observed at 1.40 K and 1.19 K, respectively. Figure 1d shows the I-V curve measured at 400 mK where a typical overdamped Josephson junction behavior is observed[41]. The corresponding dV/dI curve further validated the behavior of the Josephson junction in our heterostructure where Cooper pair tunneling is expected. The critical current of junction is equivalent to that of Al. These results suggest that compared to the $Sr_2RuO_4$, the superconductivity of Al is more sensitive to external disturbances, potentially serving as an effective electronic probe.

The main observation is shown in Figure 2a exhibiting the resistance vs. external magnetic field curves of various junctions with varying $Ar^+$ bombardment times, i.e., different disorder strength. All the junctions have the same fabrication procedures and geometry, and were cooled down to 400 mK at zero field to allow the superconducting Josephson state formed. In the junction without any $Ar^+$ bombardment (0 minute), a typical superconductor-to-normal metal transition is observed at 72 Oe. When sweeping the magnetic field back and forth between 1000 Oe and -1000 Oe, the magnetoresistive curves overlap without noticeable hysteresis. Interestingly, in junctions where $Sr_2RuO_4$ is bombarded with $Ar^+$ ions, distinct magneto-resistive hysteresis behaviors are observed. Taking the junction with 100 mins bombardment time for example, a superconductor to normal metal transition occurs (black curve) when first increasing the magnetic field. When sweeping the magnetic field back and forth, a clear butterfly shaped magneto-resistive hysteresis is observed: the red curve shows the downsweep process where the zero resistance state is seen in the negative external magnetic field region. Correspondingly, the blue curve shows the following up-sweep data where the zero resistance state appears in the positive external magnetic field region in a symmetrical manner. Here we define ΔB as the separation field between upsweep and down-sweep curves to describe the amplitude of hysteresis. Noticeably, when increasing the Ar$^+$ bombardment time from 0 to 100 minutes where the disorder on Sr$_2$RuO$_4$ surfaces are enhanced, ΔB increases monotonically and tends to saturate gradually as shown in Figure 2b. Such hysteresis behavior is persistent over time as shown in supplementary Figure S2. As a comparison, the magneto-resistive curve of a pure Al film is also measured and shown in Figure 2a. A typical field dependent superconducting to normal state transition of Al is observed without any hysteresis. Comparing the amplitude of resistance change between the Josephson junction and the pure Al, the measured superconductor to normal state transition in the Sr$_2$RuO$_4$-Al Josephson junction is governed by the Al layer, consistent with the fact that the resistance of Sr$_2$RuO$_4$ is much smaller than Al[42]. Moreover, such hysteresis behavior is reproducible under multiple field scans and observed in several samples under the same Ar$^+$ bombardment time (Supplementary Figure S3), suggesting the universality of our observation.

The magneto-resistive hysteretic behavior observed in our experiments highly resembles those in typical ferromagnetic materials with intrinsic TRSB, yet in our junctions, it is induced by disorder and linked to the superconducting natures of both Sr$_2$RuO$_4$ and Al. To understand the nature of the persistent hysteresis, we performed several control experiments to illustrate the respective roles of Sr$_2$RuO$_4$ and Al. We first demonstrate the necessity of superconducting Sr$_2$RuO$_4$ for the observed hysteresis. We prepared a junction by replacing the superconducting Sr$_2$RuO$_4$ with La-doped Sr$_2$RuO$_4$ which is not superconducting and bombarded with Ar$^+$ ions for 100 mins. As shown in Figure 2c, no hysteresis is observed in the magnetoresistive measurements. The same results are observed in Nb doped SrTiO$_3$–Al junction (Supplementary Figure S4), suggesting that the superconductivity of Sr$_2$RuO$_4$ is an essential ingredient on the observation of TRSB effect. Next we show that the superconducting Al layer serves as an essential probe to detect the disorder-induced TRSB effect in Sr$_2$RuO$_4$. Figure 2d shows the magneto-resistive curve measured at different temperatures of the Josephson junction for the 100 mins Ar$^+$ bombardment sample. With increasing temperature, the hysteresis gradually shrinks and completely disappeared over 1 K where the Al layer lost its superconductivity. The amplitude of hysteresis at different temperatures is summarized in Figure 2e. Through linear fitting, the hysteresis can reach up to 82 Oe at zero temperature. To further validate the role of superconducting Al, we replaced the Al by normal metal (Au) while preparing the Sr$_2$RuO$_4$ under the same condition. As shown in Figure 2f, the Au based junction does not present hysteresis effect. These results point towards signature of disorder-controlled TRSB in superconducting Sr$_2$RuO$_4$ detected by Al through transport measurement.

We then analyze the possible physical scenarios on the observed effect. The first possibility is the emergent ferromagnetism from the Sr$_2$RuO$_4$ surface due to stoichiometric change after Ar$^+$ bombardment, as similar magneto-resistive hysteresis is observed in some superconductor-ferromagnet hybrids[43,44]. To examine whether ferromagnetism exists on the bombarded Sr$_2$RuO$_4$ surface, we carried out magneto-optical Kerr effect (MOKE) measurements. Supplementary Figure S5 shows the field-dependent MOKE signal measured at 50 mK on the 100 min Ar$^+$ bombarded Sr$_2$RuO$_4$ surface. The Kerr signals obtained by the down-swept and the up-swept external magnetic field are linear with ($\frac{d\theta_k}{dB} \approx 0.0094 \, \mu rad/Oe$) and coincide with each other with statistical mean equal to 0.015±0.007 μrad, suggesting the absence of ferromagnetism within 1.6±0.7 Oe. Next, we exclude the granularity as the origin of our results. In granular superconductors, the superconducting state is resumed at non-zero external magnetic fields with the same sign of the starting sweeping field[45,46]. In our measurements, the minimum resistance state occurs when the external magnetic field has the opposite sign of the starting sweeping field as illustrated in Figure 2a and 2d, effectively ruling out granularity as the origin of our persist hysteresis behavior. Then, the contribution from regular bulk vortex pinning can be effectively ruled out by SQUID measurements. As shown in Supplementary Figure S6, both clean and bombarded Sr$_2$RuO$_4$ samples show similar asymmetric M-H loop indicating vortex pinning effect[47]. The fact that no hysteresis behavior is found in our transport measurement for the clean sample suggests our junction is insensitive to the regular vortex pinning. Lastly, in our junctions, as shown in Supplementary Figure S7, the hysteresis amplitude of magneto-resistive curves does not depend on the sweeping range and speed. Moreover, as shown in Supplementary Figure S8, we do not find hysteresis in the bias dependence of I-V measurement. These observations exclude the occurrence of motion of domain walls with chiral nature[48].

Having ruling out the above scenarios, we provide a plausible picture explaining our experimental observations. Regardless of the specifics, we argue that the key requirement for the observed effect is that the disorder causes a reconstruction of the superconducting state by breaking time-reversal symmetry and inducing loop current configurations. In this framework, we consider a physical scenario that can be applied to explain the hysteresis behavior of the magnetoresistance due to disorder nearby the surface of Sr$_2$RuO$_4$ in a region of the sample that is between the bombarded surface and the inner layers far from the surface. The disorder is inducing a reconstruction of the superconducting order parameter with a changeover from a time-reversal preserving pairing to a time-reversal symmetry broken configuration. Although the pairing symmetry of the Sr$_2$RuO$_4$ is not yet settled, it is known to be a multiband superconductor, and thus we consider an effective description with two bands having conventional s-wave pairing to grasp the key features of the observed effects (this can be for instance related to the $d_{xz}$ + $id_{yz}$ interorbital pairing[49]). Depending on the strength of the disorder, a reconstruction of the multiband pairing configuration can occur from 0 (i.e. $s_{++}$) to $\pi$ (i.e. $s_{+-}$) Josephson interband coupling, or to a TRSB state marked by a $s_{++} + i\,s_{+-}$ phase (Figure 3a). Hence, in the presence of an inhomogeneous spatial distribution, a loop current can be induced with a resulting flux that is trapped in the superconducting state (Figure 3b). Apart from the connection with the time-reversal symmetry breaking, the intertwinning of $\pi$ pairing and multiband electronic structure often marks the occurrence of unconventional superconducting phases, e.g. in iron-based[29,50] and oxide interface superconductors[51,52], electrically or orbitally driven superconductivity[53-57], and multiband noncentrosymmetric superconductors[51,58-60]. To describe quantitatively this scenario, we consider a two-component Ginzburg-Landau (GL) approach, where the presence of disorder is manifested through the appearance of higher order harmonics in the phase difference $\phi$ between the components of the order parameter of the GL free energy:

$$G = F_0 + \left(\tfrac{1}{2}k_{11}|\Delta_1|^2 + \tfrac{1}{2}k_{22}|\Delta_2|^2 + k_{12}|\Delta_1||\Delta_2|\cos\Phi\right)q^2 + 2(a_{12}|\Delta_1||\Delta_2| + c_{11}|\Delta_1|^3|\Delta_2| +$$
$$c_{22}|\Delta_1||\Delta_2|^3)\cos\Phi + c_{12}|\Delta_1|^2|\Delta_2|^2\cos2\Phi \qquad (1)$$

where $F_0$ is the homogeneous part of the free-energy (see Supplemental Material [39]), while the wavevector $q$ allows to describe the formation of a spontaneous loop current state with a non-trivial interband phase relation $\phi^*$. Indeed, the wave-vector $q$ is proportional to the magnetic field and to the intrinsic winding number $N = \pm 1$ of the induced loop current states (see Figure 3a,b), which is related to the direction of the magnetic field[61]. The phenomenological coefficients of the GL in Eq. (1) can be derived from the microscopic Usadel equations[10]. It is important to note that $c_{ij}$ and $k_{12}$ are absent in the case of a clean two-component superconductor. They are a direct consequence of the contribution of the interband disorder, whose strength is characterized by the microscopic parameter known as the interband scattering rate $\Gamma$, being proportional to the impurity concentration[1]. Hence, one can observe that depending on the magnitude of the applied magnetic field, a switching between TRS and TRSB states can occur. Figure 3c illustrates this process from the energetic point of view. The red and blue curve correspond to the ground state of the system calculated on the basis of Eqs. (1) and on the values of the minima of the GL with different winding of the loop current state (i.e. $N = -1$ (red curve) and $N = 1$ (blue line)). Hence, one can argue that the increase of the disorder strength due to the Argon bombardment corresponds to the occurrence of doubly degenerate ground state in the multicomponent superconducting system as shown in Figure 3c. Increasing the disorder strength leads to a larger difference in energy between loop current states with opposite winding. This implies that one needs a greater magnetic field to have a configuration with equal concentration of opposite winding loop states. The linear dependence of the magnetization can be accounted by the fact that phase reconstruction of the superconducting state is not depending on the strength of the superconducting gap and thus we do not have an activation mechanism.

Given the above physical picture, the superconducting to normal state transition in the Al layer is an essential ingredient to probe the trapped flux from the reconstructed superconducting state in Sr$_2$RuO$_4$. When the external magnetic field is reduced from positive values to zero, the trapped flux acts as effective magnetic field so that the Al layer near the interface remains in the normal state thanks to its low critical field. When the external magnetic field is negative to partially cancel the trapped flux in Sr$_2$RuO$_4$, the Al layer will return to the superconducting state to allow the cooper pair tunneling so that the zero-resistance state is resumed. Further sweeping towards the negative magnetic field will drive the Al back to the normal state and in the mean time, reverse the winding number of the loop current. The same process occurs when the external magnetic field is swept from negative to positive, causing a sizable magneto-resistive hysteresis observed in experiment.

In conclusion, our results provide evidences of disorder induced TRSB effects which are directly probed by magneto transport variations within a Josephson junction setup. This result points to an order parameter reconstruction nearby the surface of the examined superconductor. Our findings present innovative Josephson toolkit for the electronic identification of TRSB effects, which can be potentially relevant for superconducting electronics too.

**Conflict of interest**

The authors declare that they have no conflict of interest.

**Author Contributions**

Y.W., M.C., H.G., and J.S. initiated the project. A.G., R.F, A. V. grow the samples, Y.W. performed devices fabrication and transport measurements, D.H., Y.L. performed MOKE measurements, C.M., Z.W. performed STEM measurements. Y.Y., M.C. performed theoretical calculations. All authors discussed the results.

**Acknowledgments**

This work is primarily supported by National Key Research Program of China (2022YFA1403300, 2020YFA0309100), National Natural Science Foundation of China (12074073, 12204107), Shanghai Municipal Science and Technology Major Project (2019SHZDZX01) Shanghai Municipal Natural Science Foundation (22ZR1408100, 23ZR1407200) Shanghai Science and Technology Committee (21JC1406200). M.C., R.F., and A.V. acknowledge support from the EU's Horizon 2020 research and innovation program under Grant Agreement No. 964398 (SUPERGATE). M.C. acknowledges financial support from PNRR MUR project PE0000023-NQSTI. M.C., R.F., A.G., and A.V. acknowledge partial support by the Italian Ministry of Foreign Affairs and International Cooperation, grant number KR23GR06. Part of the experimental work was carried out in the Fudan Nanofabrication Laboratory. We thank fruitful discussion with Pro. Cheng Zhang, Prof. Yihua Wang, Prof. Yang Qi.

**Appendix A. Supplementary materials**

Supplementary materials to this article are attached in a separate file.

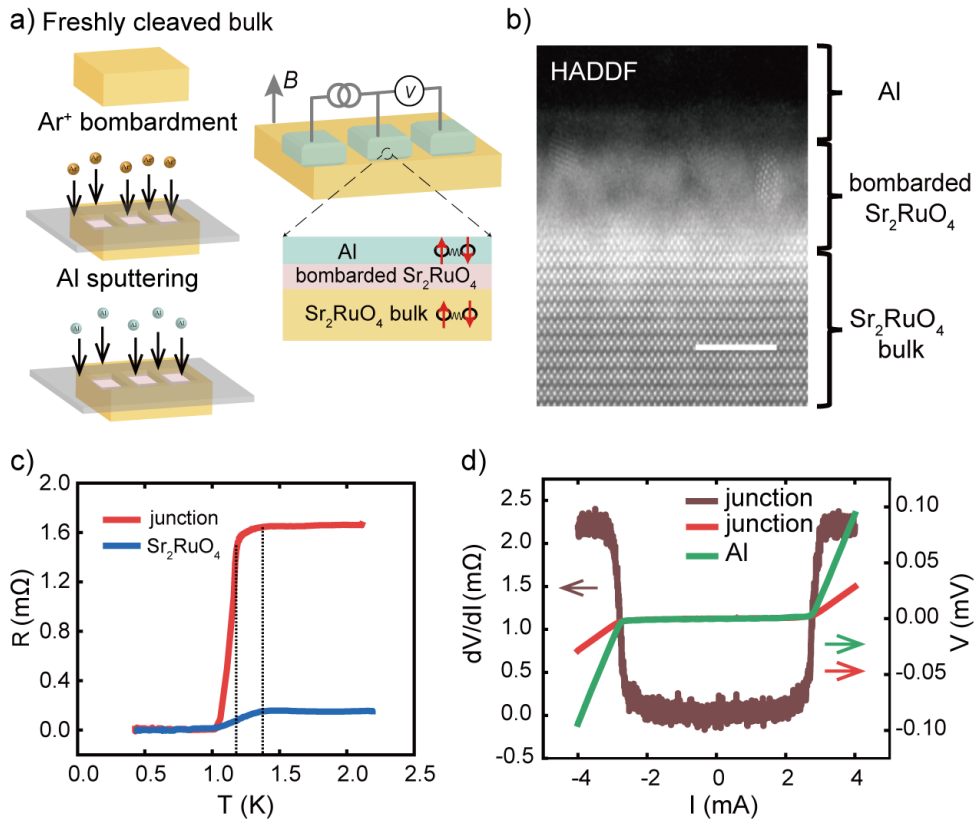

FIG. 1. Fabrication and basic electronic properties of Josephson junction. a) Schematics on the preparation process and measurement geometry. b) cross-sectional STEM image of junction with 100 min Ar+ bombardment time on the Sr2RuO4 surface. Scale bar is 5nm c) Resistance vs temperature curves of Sr2RuO4-Al junction in b) (red) and pure Sr2RuO4 bulk (blue). d) The corresponding dV/dI (brown) and I-V curves of Sr2RuO4-Al junction (red), and I-V curve of Al film (green) measured at T=400 mK.

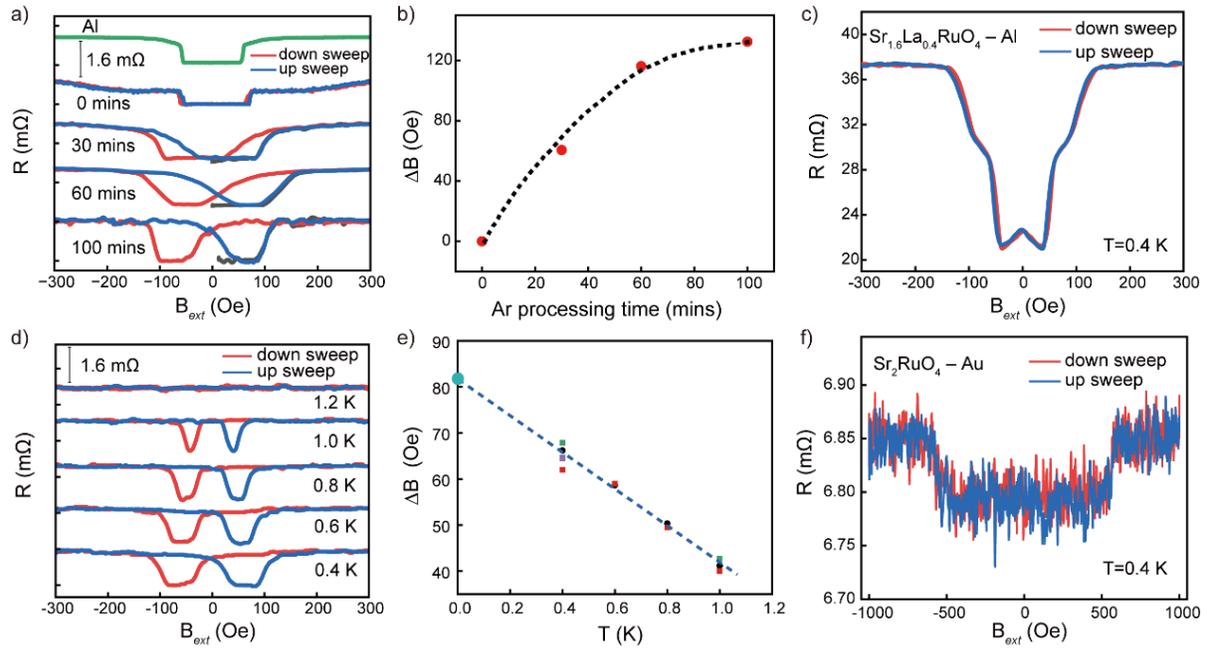

FIG. 2. Experimental signature on disorder-induced TRSB. a) Magneto-resistive curves of Josephson junctions with varying Ar+ bombardment time from 0 to 100 minutes and pure Al (green). Data were obtained at T=400 mK and vertically shifted for 1.6 mΩ for clarity. b) The amplitude of hysteresis as function of the Ar+ processing time (dotted line for eye guidance). c) Magneto-resistive curve for $Sr_{1.6}La_{0.4}RuO_4$-Al junction. d) Magneto-resistive curves of $Sr_2RuO_4$-Al Josephson junction at various temperatures. Hysteresis disappears at 1.2 K where Al film is at normal state. e) Linear fitting on the magnitude of hysteresis with temperature. Different colors represent the values obtained from various junctions. f) Magneto-resistive curve for $Sr_2RuO_4$-Au junction.

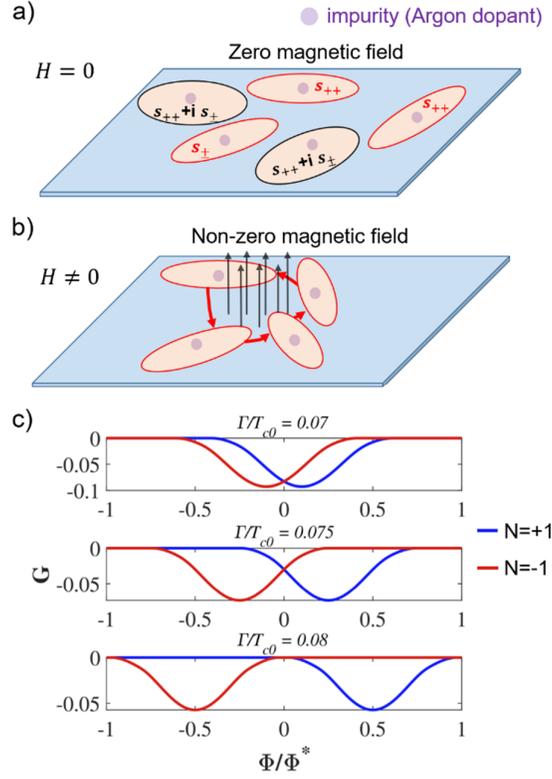

FIG. 3. a) Sketch of the disorder induced superconducting reconstruction for an effective two-band superconductor showing the formation of π-pairing (i.e. $s_\pm$) and TRSB state (i.e. $s_{++} + is_\pm$) around the defects. The presence of an external magnetic field can lead to superconducting loop current states with spontaneous magnetic fluxes as schematically depicted in (b). In panel (c) we show the evolution of the free energy as a function of the effective flux ($\Phi/\Phi^*$) generated by the superconducting loop currents for different values of the disorder strength and assuming clockwise (N = 1) and anticlockwise (N = -1) current flow. The value of $\Phi^*$ is of the order of hundreds Oe assuming a characteristic length of the loop current state of few coherence length in the ab plane. The minimum of the free energy is for a nonvanishing flux that grows with the increase of the disorder strength. We notice that the energy separation between the opposite winding configurations also increases with the disorder strength. This trend provides evidence of a stronger TRSB effect as a function of the disorder.